\documentstyle[twocolumn,prb,aps]{revtex}
\begin{document}

\draft
\title{Electronic density of states derived from thermodynamic critical field curves 
for underdoped $La_{2-x}Sr_xCuO_{4+\delta }$}

\author{Yung M. Huh and D. K. Finnemore}
\address{Ames Laboratory, USDOE and Department of Physics and Astronomy,\\
Iowa State University, Ames, IA 50011}

\date{\today}
\maketitle
\begin{abstract}
Thermodynamic critical field curves  have been measured for 
 $La_{2-x}Sr_{x}CuO_{4+\delta }$ over the full range of carrier concentrations  
where superconductivity occurs in order to determine changes in the normal state density of 
states with carrier concentration.  There is a substantial window 
in the $H-T$ plane where the measurements are possible because the samples are both thermodynamically 
reversible and the temperature is low enough that 
vortex fluctuations are not important.  In this window, 
the data fit Hao-Clem rather well, so 
this model is used to determine $H_c$ and $\kappa _c$ for each temperature and 
carrier concentration.  
Using $N(0)$ and the ratio of the energy gap to transition temperature, $\Delta (0)/k_BT_c$, 
as fitting parameters, the $H_c~vs~T$ curves give $\Delta (0)/k_BT_c \sim  2.0$ over 
the whole range of $x$.  Values of $N(0)$ remain rather constant in the 
optimum-doped and overdoped regime, but drops quickly toward zero in the underdoped 
regime. 
.

\end{abstract}


\section{Introduction}

In the study of superconductivity in the cuprate materials, the density of states at the Fermi surface, $N(0)$, is a key variable in the determination 
of the transition temperature.  The $La_{2-x}Sr_{x}CuO_{4+\delta }$ system 
is a doped insulator for $x=0$ and the carrier concentration can be increased 
in an orderly way by holding the oxygen concentration constant and increasing 
the Sr content.\cite {1}  Numerous specific heat measurements\cite {2,3} have been carried out to gain insight into the changes in the density of states.   The determination of $N(0)$ 
from specific heat data, however, is made difficult because the lattice specific heat 
is large and because the upper critical field is above the reach of the 
magnets available.  Another approach to the changes in $N(0)$ is through the 
study of optical conductivity.\cite{4}  By assuming a Drude like absorption 
and integrating over all frequencies, it was found that $N(0)$ rises in the 
underdoped region and levels off in the overdoped region.

The purpose of this work is to measure the thermodynamic critical 
field curves, $H_c~vs.~T$, for the $La_{2-x}Sr_{x}CuO_{4+\delta }$
system over the entire range of superconductivity.
Thermodynamic critical field data are important in their own
right because they give the free energy difference
between the superconducting and normal state and a
good starting place for theory.  In addition, we wish to compare the thermodynamic 
critical fields of   $La_{2-x}Sr_{x}CuO_{4+\delta }$ with those of classical superconductors  and with the Bardeen-Cooper-Schrieffer (BCS) picture.\cite {5}  Within the BCS model and succeeding strong coupling 
modifications,\cite {6}  the ratio of the critical field at $T=0$ to the critical temperature,  $H_o/k_BT_c$ is related
to $N(0)$ and the curvature of the $H_c~vs.~T$ curve is related
to the ratio of the energy gap at $T=0$ 
to transition temperature, $\Delta (0)/k_BT_c$.  The goal is to measure 
these changes in $N(0)$ and  $\Delta (0)/k_BT_c$ as the $Sr$ content changes.

There is a vocabulary problem in the discussion of these data in that the thermodynamic 
critical field curve and the free energy difference are most closely related to the 
condensation energy per Cooper pair rather than the energy gap.  In BCS, the two quantities  are equivalent, and we have used the terms interchangeably in the discussion here because we are comparing the results to BCS.  
 
A preliminary single crystal study recently published\cite {7} illustrates the success 
and difficulties of measuring thermodynamic critical fields in this system.        
Extensive measurements\cite {7} of a high quality single crystal of
$La_{1.90}Sr_{0.10}CuO_{4+\delta }$ have shown that the 
superconducting magnetization curves,
$M_{sc}~vs.~H$, 
of this moderately underdoped high temperature superconductor obeys the 
Hao-Clem model\cite {8} over a wide range of temperature, $T$,
and magnetic fields, $H$.
The Hao-Clem model is a variational calculation developed originally for the 
$YBa_2Cu_3O_{7-\delta }$ system and it was shown to describe the magnetization curves,   well for both 
$YBa_2Cu_3O_{7-\delta }$\cite {9} and $La_{2-x}Sr_{x}CuO_{4+\delta }$ 
system.\cite {7}  In the $La_{1.90}Sr_{0.10}CuO_{4+\delta }$ studies, it also 
was found that the normal state background magnetization
follows a 2D Heisenberg model\cite {10} for the $Cu$
spins with values close to those measured
previously by Nakano and coworkers.\cite {11}
In addition, this $x=0.10$
sample showed a vortex fluctuation contribution to the magnetization\cite {12} 
 close to $T_c$.  The presence of these fluctuations limits the temperature range 
of applicability of Hao-Clem because they introduce a new term in the free energy.  We 
only use Hao-Clem in regions where these fluctuations do not contribute significantly.  
At low temperature  irreversibility of the
 magnetization  prevents equilibrium magnetization measurements and close
 to $T_c$, fluctuations make Hao-Clem inapplicable.  Even so, there is
 a window for each x-value where Hao-Clem works well and thermodynamic
 critical fields can be measured.

\section{Experiment}
Both single crystals and grain-aligned-powders are used in this work.
In the
early part of the experiment, it was thought that it would be easier
to control the oxygen content in grain-aligned-powders than in single crystals 
because the diffusion distance for oxygen would be less.  Single crystals
from
three different sources, however,  have given essentially the same result as the powders.  
Magnetically aligned samples were prepared by grinding appropriate amounts 
of $La_2O_3$, $SrCO_3$ and $CuO$ in an agate mortar and pestle.  Mixed 
and ground powders were pressed into pellets, placed in an alumina boat, and 
initially fired for $24 h$ at $750^{\circ }C$.  This was followed by
grinding,
pelletizing, and firing several times at $850^{\circ }$C
and $970^{\circ }$C for
$48$ and $72 h$ respectively.  After several cycles at these temperatures,
the
grinding, pelletizing and sintering was done at successively higher temperatures
of $1000 ^{\circ }$C, $1050^{\circ }$C, and $1100^{\circ }$C
in a tube with oxygen flowing at $2.5 cm^3/min$
for $24 h$ each time.  Measurements
of $T_c$ and zero-field-cooled Meissner screening fraction at $1.0~mT$ were 
carried out as the first test for sample quality.  The final pellet was ground 
to a particle size of about $20~\mu m$, mixed with a low viscosity epoxy 
(Epotek 301), oriented in a magnetic field of $8.0~T$, and then the epoxy 
was allowed to harden.  X-ray diffraction patterns showed only $(00\ell )$
peaks.  The full width at half maximum of the $(008)$ peak was
$5^{\circ }$.  Inductively coupled plasma (ICP) measurements
were used to determine the $Sr$
content or x-value.

Magnetization data were taken with $H\parallel c$ in a Quantum Designs magnetometer with a $3-cm$ scan for the $x=0.13$ single crystal and a $6-cm$ 
scan for all others.  In this magnetometer, the sample is moved through a series of three pick-up 
coils, and the length of scan determines the uniformity of magnetic field seen by the sample  during the scan.  We selected the sample that has the highest $T_c$ for 
any given x-value and this corresponded well with literature values in all 
cases.  In the region of $x=1/8$, the superconducting transition widths are 
about double the width away from this region.  In this work, the magnetization 
is taken to be thermodynamically reversible if the difference in the field increasing 
and field decreasing magnetization is less than one percent of the average magnetization.   
The range of temperatures was $1.4~K$ to $300~K$, and the range of magnetic fields was 
$0$ to $7~T$.

\section{Results and discussion}

The superconducting transition temperature, $T_c$, for both the single 
crystals (solid triangles) and the grain aligned powders (solid circles) are 
rather close to the literature values as shown in Fig. 1.  
There is a very clear dip in the curve 
near $x=1/8$ similar to that found by others and
in $La_{2-x}Ba_{x}CuO_{4+\delta }$.\cite {13}  The $T_c$ values for the grain
aligned samples agree well with the single crystals published here as well as 
those reported in the literature.\cite {14}

\subsection{Normal State Magnetization}

There is a very predictable and regular background magnetization, $M_b$,
that
appears to arise primarily from the Cu spin system.  Measurements of $M_b$
in the normal state were made between $60~K$ and $200~K$ and found
to fit the relation, 

$$M_b~=~CH~+~M_s~tanh~(\beta H)~~,  \eqno (1)$$  

where the dominant term is the linear $CH$ term.  In Eq.1, $C$, $M_s$, and $\beta $ 
are fit constants.
Figure 2 shows that the magnitude 
of this magnetization is on the order of $10^{-5}~to~10^{-3}\mu _B/Cu~atom$. 
As reported earlier for the $x=0.10$ single 
crystal, the $M_s~tanh~(\beta H)$ term saturates at about $0.1~T$ and the 
value of $M_s$ ranges between $3-6\times 10^{-5}\mu _B/Cu~atom$.  The 
$M_s~tanh~(\beta H)$ term is small and both $M_s$ and $\beta $ are nearly
independent of temperature
and x-value in the range measured.  The value of the slope of the linear
term,
C, is rather close to the values reported by Nakano and coworkers\cite {11} 
as shown by Fig. 3.  The solid lines are from Ref. 5 for 
$x=0.10, 0.14, 0.16, 0.20, 0.26$
running from bottom to top.
To determine the normal state magnetization at temperatures below $T_c$,
the
value of $C$ is linearly extrapolated from the $60~K$ to $200~K$ data to the 
desired temperature and $M_b$ is determined from Eq. 1.
Because the constants, $M_s$ and $\beta $ do not change with temperature,
they are
 taken to be the average of the values determined from $60-200~K$.

\subsection{Irreversibility field}

In the over-doped region, the irreversibility field, $H_{irr}$, has 
 essentially the same values as the optimum doped sample as shown in the reduced temperature, $T/T_c$ plot of Fig. 4.  In the underdoped region, the flux pinning is much weaker and $H_{irr}$ drops to lower values.  As Janossy and coworkers\cite {15} showed for 
$YBa_2Cu_3O_{7-\delta }$, as charge carriers are removed from the cuprate 
superconductor, the material becomes more anisotropic and the pinning is 
reduced.  

\subsection{Thermodynamic critical field curves}

To determine the magnetization of the superconducting charge carriers,
 $M_{sc}$, the background magnetization  is subtracted from the 
total measured magnetization, $M_t$, by, $M_{sc}=M_t-M_b$.
To illustrate the results, Fig. 5 shows superconducting
magnetization data as a function of magnetic field for the interval
between $H_{irr}$ and the field where vortex
fluctuations become important.  In the first attempt to fit these data to 
the Hao-Clem model, both $H_c$ and $\kappa _c$ are used as adjustable 
parameters.  Here, $\kappa _c = H_{c2}/\sqrt 2 H_c$ is approximately the 
ratio of the penetration depth to the coherence distance.  
Usually the data do not extend very close to $H_{c2}$, so  
the fit is not very sensitive to $\kappa _c$.  Hence, the average of $\kappa _c$ 
for any given $x$-value is 
used and $H_c$ is adjusted slightly for the best fit.  For the optimum-doped and 
overdoped range, $\kappa _c$ ranges from 80 to 120.  The maximum value of 
$\kappa _c$ is about $200$ for $x=0.10$ and it falls to about $120$ again 
in the $x$ equal to $0.07-0.08$ range.  The fits to
Hao-Clem and the resulting $H_cvs.T$ curve are shown in Fig.6.
     
The critical field curve data for samples in the overdoped
regime are shown
in Fig. 7a, and the data for the underdoped regime are shown in Fig. 7b.
The
solid lines through the data are best fit parabolas of the
form  $H_c=H_o[1-(T/T_c)^2]$.  In the overdoped data of Fig. 7a,
the ratio of
$H_o/T_c$ remains relatively constant,
and in the underdoped regime, the ratio
of $H_o/T_c$ drops quickly as the x-value or charge carrier concentration 
is reduced.

\subsection{Comparison of data with BCS theory}

The BCS theory\cite {5} has been very successful 
describing the thermodynamic critical field curves of classical 
superconductors.  Within the theory, the  
$H_c$ value at any given temperature can be calculated from two microscopic variables, 
the ratio of the energy 
gap to the thermal energy, $\Delta (T)/k_BT$ at that temperature and the 
value of the density of states, $N(0)$.
Modifications of BCS to include strong-coupling effects\cite {6} show that
the ratio
of the energy gap at $T=0$ to $k_BT_c$, $\Delta (0)/k_BT_c$
rises above the
weak-coupling BCS value of $\Delta (0)/k_BT_c=1.76$ to values above 
$\Delta (0)/k_BT_c=2$ for strong-coupling superconductors like $Pb$ and $Hg$.  For these materials, the temperature dependence of $\Delta (T)/\Delta (0)$ 
has approximately the same shape as the original BCS form.    

To compare and contrast these $La_{2-x}Sr_xCuO_{4+\delta }$ 
superconductors with classical superconductors, we assume that 
strong-coupling BCS theory applies to 
both classes of material and do a two parameter fit of the
thermodynamic critical field curves using $N(0)$ and
$\Delta (0)/k_BT_c$ as fitting
variables.  In this fitting, $N(0)$ controls the magnitude of $H_c$ and 
$\Delta (0)/k_BT_c$ controls the curvature of the $H_cvs.T$ plot.  
The results are shown in Fig. 8a and 8b.  The value of $\Delta (0)/k_BT_c$ 
is consistently in the $1.9$ to $2.1$ range with the underdoped samples being 
very similar to the overdoped samples.  The specific heat $\gamma $, 
which is related to $N(0)$ by $\gamma =2/3\pi ^2k_B^2N(0)$, is relatively 
constant at $3.0~mJ/mol~K^2$ in the optimum and overdoped region.  The value 
of $\gamma $ then drops quickly to a minimum at the $x=1/8$ region,
rises to a
maximum at $x=0.010$, and falls quickly as $x$ falls in the underdoped
regime.  Measured values of $T_c$ are also shown in Fig. 8b for
comparison with the
trajectory of $\gamma ~vs.~x$. A much more detailed discussion of the
data and the analysis is given elsewhere.\cite {16}

\section{Conclusions}

There is a reasonable window of thermodynamic reversibility where equilibrium magnetization curves can be measured in the 
$La_{2-x}Sr_{x}CuO_{4+\delta }$ system. The superconducting magnetization 
curves obey Hao-Clem rather well, even in the underdoped region as long as 
data are taken in the range of applicability of the model.
As the x-value drops, $H_{irr}$ drops giving a wider region
of the $H-T$ plane. The
range of vortex fluctuations, however, also increases
so that the temperature interval where Hao-Clem can be used remains
roughly constant
over the x-values measured.

Thermodynamic critical field curves for $La_{2-x}Sr_xCuO_{4+\delta }$ are 
rather similar to those of classical superconductors.  From the curvature of 
the $H_cvs.T$ plots, one can derive a $\Delta (0)/k_BT_c$ value that is 
roughly independent of $x$ and equal to about 2.0.
From the magnitude of the
$H_cvs.T$ plot, the specific heat $\gamma $ remains nearly constant
at $3.0~mJ/mol~K^2$ in the overdoped region.  In the underdoped region,
$\gamma $ falls to a minimum at the $x=1/8$ value, rises to a maximum at 
$x=0.10$, and then falls quickly toward zero as $x$ approaches $0.06$.  
In the broad picture, $N(0)$ is roughly constant in the overdoped regime, 
and $N(0)$ falls quickly in the underdoped regime.  In this broad feature, 
the data are similar to the change in the density of states found by integrating 
optical absorption data over all frequencies.\cite {4}

\section{Acknowledgments}

Ames Laboratory is operated for the
U. S. Department of Energy by Iowa State University under contract No.
W-7405-ENG-82 and
supported by the DOE, the Office of Basic Energy Sciences.

\begin{figure}
\caption{$T_c$ values for both single crystals (solid triangles) and grain aligned powders 
(solid circles)}.  The solid line is a sketch from data of Ref. 14.
\end{figure}

\begin{figure}
\caption{Normal state magnetization at $80~K$ for the full range of x-values.  The solid lines are 
fits of the data to $Eq. 1$.  The inset expands the low field portion.} 
\end{figure}

\begin{figure}
\caption{Comparison of the slope of the normal state magnetization with 
the values obtained by Ref. 5.  Bottom to top, the solid lines for Ref. 5 
run from $x=0.10, 0.14, 0.16, 0.20, 0.26$ }
\end{figure}

\begin{figure}
\caption{Irreversibility curves.}
\end{figure}

\begin{figure} 
\caption{$M_{sc}vs.H$ for $x=0.156$ grain aligned sample.}
\end{figure}

\begin{figure}
\caption{A fit of the grain aligned $x=0.156$ data to the universal Hao-Clem curve for $\kappa _c = 91$.  
The inset shows the value of $H_c$ used to fit the curves.}
\end{figure}

\begin{figure}
\caption{Thermodynamic critical field curves in both a) overdoped region, 
and b)optimum doped and underdoped regions.}
\end{figure}

\begin{figure}
\caption{Results of a two parameter fit to $H_c~vs.~T$ data, 
a)$\Delta (0)/k_BT_c$, and b) specific heat $\gamma $.}
\end{figure}

\vfil\eject

\end{document}